Article

# Skyrmion lattice domain formation in a non-flat energy landscape

Check for updates

Raphael Gruber[1], Simon M. Fröhlich[1], Jan Rothörl[1], Maarten A. Brems[1], Tobias Sparmann[1], Fabian Kammerbauer[1], Maria-Andromachi Syskaki[1,2], Elizabeth M. Jefremovas[1], Sachin Krishnia[1], Asle Sudbø[3], Peter Virnau[1] & Mathias Kläui[1,3]

Magnetic skyrmions are chiral spin structures with non-trivial topology that comprise two-dimensional quasi-particles and are promising information carriers for data storage and processing devices. Skyrmion lattices in magnetic thin films exhibit Kosterlitz-Thouless-Halperin-Nelson-Young (KTHNY) phase transitions and have garnered significant interest for studying emergent 2D phase behavior. In experimental skyrmion lattices, the main factor limiting the quasi-long-range order in thin films has been the non-flat energy landscape – often referred to as pinning effects. We demonstrate direct control of the skyrmion lattice order by effectively tuning the energy landscape employing magnetic field oscillations. By quantifying lattice order and dynamics, we explore how domain boundaries form and evolve due to pinning effects in Kerr microscopy experiments and in Brownian dynamics simulations, offering a pathway to control and study emergent skyrmion lattice properties and 2D phase behavior.

Two-dimensional (2D) lattices encompass a unique nature of ordering phenomena[1–5], qualitatively different from 3D or other dimensions. Specifically, the transition between the solid phase—where translational quasi-long-range order (QLRO) is present—and the isotropic liquid phase can be marked by the emergence of an intermediate hexatic phase[3,4]. This hexatic phase possesses only orientational, but no translational QLRO and is described within the framework of the Kosterlitz–Thouless–Halperin–Nelson–Young (KTHNY) theory[1–5].

Magnetic skyrmions—chiral spin textures that exhibit quasi-particle properties due to their topologically non-trivial spin structure[6–8]—are an ideal platform for exploring the fundamental physics of ordering in 2D[9–13]: In thin magnetic films with layer thicknesses of around 1 nm, skyrmion sizes can range from nanometers to a few micrometers[8]. This size range, combined with the uniformity of their texture across the film thickness, makes those skyrmions ideal 2D quasi-particles[12]. In Ta/CoFeB/MgO multilayer stacks, the repulsive interaction potential[14–17] enables skyrmion arrangements of high densities to form ordered hexagonal lattices at room temperature[10,12,16]. Thereby, the thermal excitation of spins is sufficient to cause Brownian motion of the quasi-particle skyrmions[18–20]. The skyrmion dynamics is even tunable on the fly[21] and the individual skyrmions can be tracked with high resolution imaging by Kerr microscopy in real-time and -space[18–20]. Therefore, their versatility in order to induce and observe the dynamics associated with the KTHNY phase transitions[10] is a key advantage of skyrmions over previously studied colloidal[22,23] or superconducting vortex[24,25] systems.

However, achieving QLRO for skyrmion lattices in continuous thin films is challenging because of the underlying non-flat energy landscape[10–12,26]. The roughness of the energy landscape arises from non-uniform magnetic properties[20,27] due to non-uniform interfaces or locally varying crystallinity within the material stack[20,27–31]. The material inhomogeneities cause preferred positions for the skyrmions – so-called pinning sites. The term *pinning* is often used to describe the general roughness of the energy landscape. It is however important to note that the overall non-flat landscape may feature both attractive and repulsive positions[20,27–31] of varying strengths as a continuous variation of the energy of a skyrmion across the sample. In the energy landscape of thin films, skyrmions are typically pinned at their delineating domain wall, which makes pinning effects dependent on the skyrmion size[20] and effectively tunable by magnetic field oscillations[21]. While theoretical predictions allow for the existence of QLRO in skyrmion lattices with weak pinning, stronger pinning suppresses ordering transitions entirely[27,32,33]. Experimental investigations have revealed that the non-flat energy landscape significantly impacts skyrmion lattice formation[11,12], resulting in polycrystalline lattice domains, which are separated by domain boundaries (DB). Hence, QLRO is maintained only on a local scale within a domain, but broken by the DBs[11,12]. Overcoming the non-flat energy is therefore the key challenge to achieve single-crystal

[1]Institute of Physics, Johannes Gutenberg-Universität Mainz, Mainz, Germany. [2]Singulus Technologies AG, Kahl am Main, Germany. [3]Center for Quantum Spintronics, Department of Physics, Norwegian University of Science and Technology, Trondheim, Norway. e-mail: klaeui@uni-mainz.de

skyrmion lattices and to study emergent 2D phases and phase transitions on large scales.

In this article, we enhance the size of polycrystalline skyrmion lattice domains by exploiting the effective reduction of pinning due to magnetic field oscillations[21]. We study the interplay between the non-flat energy landscape and the lattice order. Specifically, we show that skyrmions, which stay pinned due to the effects of the non-flat energy landscape, cause domain boundaries between the lattice domains and support our experimental findings with Brownian dynamics simulations. Since enhancing lattice domain sizes towards true QLRO is the main challenge to observe true 2D phase behavior of skyrmions in thin films[10–12], our results pave the way to experimentally explore the statics and dynamics of 2D phases and phase transitions in 2D skyrmion lattices.

## Results and discussion

### Field oscillations assist lattice ordering

We use a Ta(5 nm)/$Co_{20}Fe_{60}B_{20}$(0.9 nm)/Ta(0.08 nm)/MgO(2 nm)/$HfO_2$(3 nm) magnetic multilayer stack to nucleate densely packed skyrmions and establish magnetic out-of-plane (OOP) contrast in real space at 16 frames per second using Kerr microscopy (see Methods section for details). After nucleation, the field of view of $200 \times 150\,\mu m^2$ contains between 4000 and 5000 skyrmions. Using the *trackpy* Python package[34], we detect the skyrmions and calculate the local order parameter

$$\psi_6\left(\mathbf{r}_{\mathrm{j}}\right) = \frac{1}{N_{nn}} \sum_{k=1}^{N_{nn}} e^{-i6\theta_{\mathrm{jk}}}$$

applying a Voronoi tessellation[35] to determine the nearest lattice neighbors. Here, $\mathbf{r}_{\mathrm{j}}$ denotes the position of a particle with $N_{nn}$ nearest neighbors at $\mathbf{r}_{\mathrm{k}}$, and $\theta_{\mathrm{jk}}$ represents the angle of the connecting vector $\mathbf{r}_{\mathrm{k}}$-$\mathbf{r}_{\mathrm{j}}$ relative to a fixed arbitrary axis[3]. The phase of the complex value defines the local orientation $\alpha = \arg(\psi_6)/6$ of the lattice.

Figure 1a shows the positions of the skyrmions as nucleated ($t_0 = 0$ s), with their local orientations $\alpha$ color-coded. The schematics next to the color bar illustrates examples of different $\alpha$, where the lattice orientation is highlighted by the differently aligned crystal axes depicted as gray lines. Note that the hexagonal lattice is six-fold symmetric – i.e., the color bar is cyclic with $\alpha = -30°$ being equivalent to $\alpha = 30°$. Figure 1b shows the same skyrmion lattice, but after waiting for $t_1 = 60$ s while an oscillating field of $A = 180$ μT peak-to-peak amplitude at a frequency $f = 100$ Hz is applied. Both at $t_0$ and $t_1$, we find a multi-domain structure with regions of equivalent orientation $\alpha$, which we identify as lattice domains. Between those lattice domains, $\alpha$ changes abruptly, forming a domain boundary (DB) which breaks QLRO. The inset shows the polar Kerr microscopy snapshot of the corresponding region, where every white dot represents one skyrmion. During the waiting time of 60 s, we find a significant increase of the lattice domain size.

To quantify the lattice domain size, we calculate the orientational correlation function

$$G_6(r = |\mathbf{r}_{\mathrm{j}} - \mathbf{r}_{\mathrm{k}}|) = \left\langle \psi_6^*(\mathbf{r}_{\mathrm{j}})\psi_6(\mathbf{r}_{\mathrm{k}}) \right\rangle \propto \exp\left(-r/\xi_6\right)$$

as spatial quantifier of orientational order. In the absence of QLRO, $G_6$ decays exponentially with a correlation length $\xi_6$, which provides a measure for the lattice domain size[3,4,10]. While the lattice domains also grow slightly at constant applied field, the field oscillations significantly enhance the ordering effect. In Fig. 1c, we demonstrate that introducing field oscillations ($f = 100$ Hz) can significantly enhance both the orientational order $\langle|\psi_6|\rangle$ and the domain size measured by $\xi_6$. As field oscillations effectively reduce pinning[21], skyrmions are enabled to rearrange into larger lattice domains. Similar driving mechanisms have already been reported to lead to depinning and lattice formation of superconducting vortices[36]. In principle, alternative driving mechanisms like current-induced motion[26,37], local heating[38] or surface acoustic waves[39] have also been shown to assist skyrmion depinning and therefore provide potential alternatives to stabilize skyrmion lattice

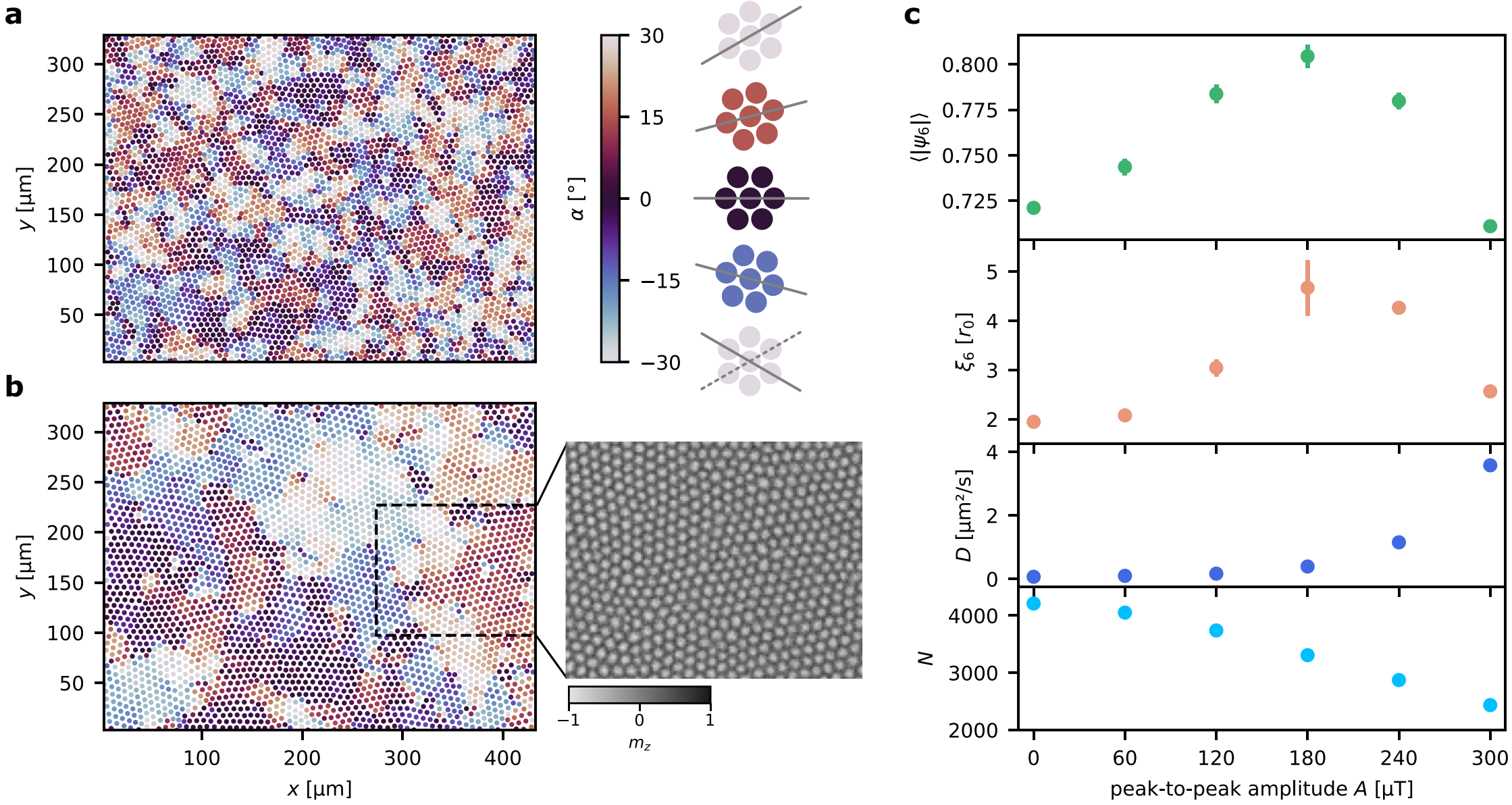

**Fig. 1 | Skyrmion lattice domain growth in oscillating fields. a** Skyrmion positions colored by the lattice orientation $\alpha$ directly after nucleation. The orientation $\alpha$ (illustrated as gray line in color-bar label) is determined by the neighbor positions and is 6-fold symmetric. **b** After 60 s in an oscillating magnetic field (180 μT peak-to-peak at 100 Hz), the lattice domains have grown significantly. The insets show the corresponding Kerr microscopy image with greyscale representing the OOP-component $m_z$ of the magnetization. **c** Influence of the field oscillation amplitude on local order parameter $\psi_6$, orientational correlation length $\xi_6$, skyrmion diffusion coefficient $D$ and number of skyrmions $N$. Data points and error bars denote average and standard deviation, respectively, of 10 s window (160 snapshots between 50 and 60 s after nucleation), averaged for three independent measurements each. As a trade-off of effectively reduced pinning and increased random diffusion, $\psi_6$ and $\xi_6$ (as measures of the order and lattice domain size, respectively) peak around $A = 180$ μT. Generally, increased oscillation amplitudes enhance the diffusion $D$ and reduce the number of skyrmions $N$.

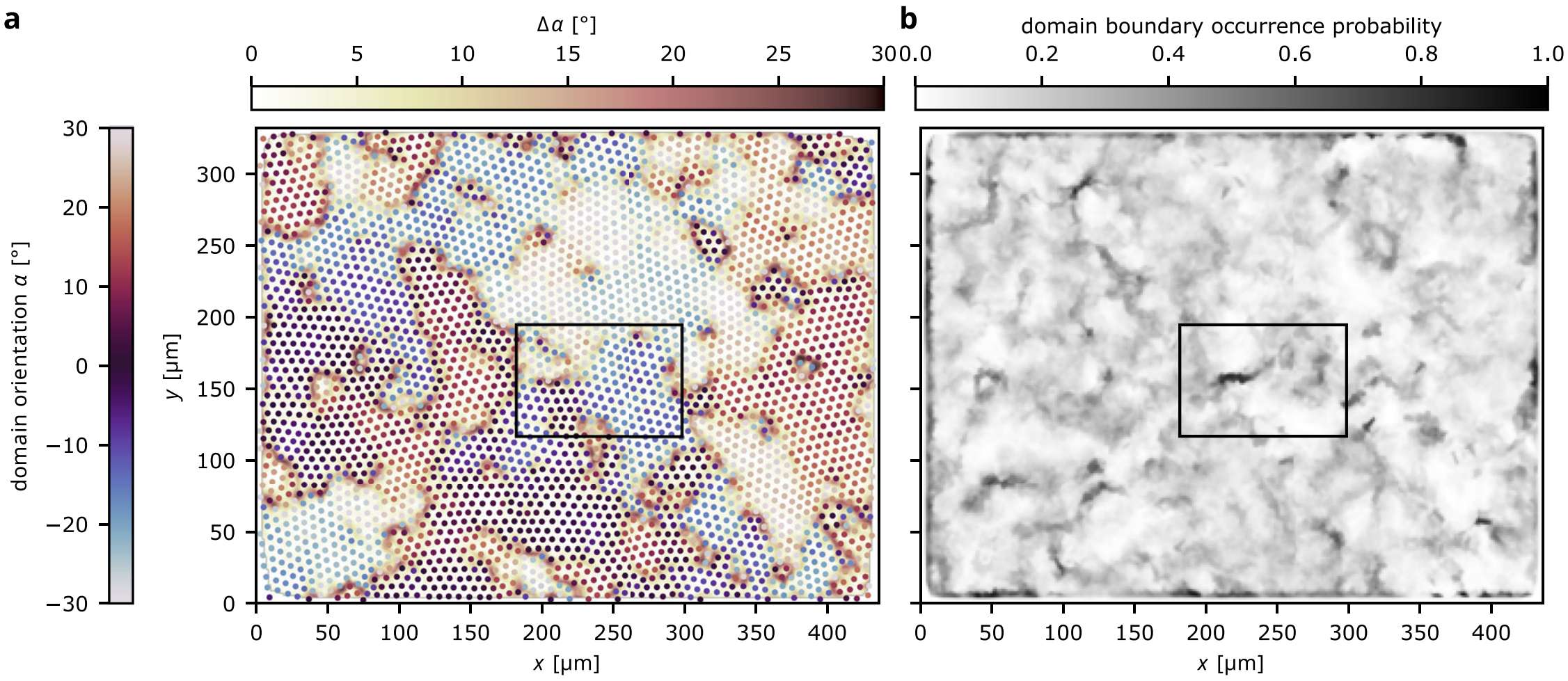

**Fig. 2 | Lattice domain boundary analysis. a** The scattered dots mark the skyrmion positions and local orientation $\alpha$ as in Fig. 1b. The background colormap visualizes the absolute change $\Delta\alpha$ in orientation between neighboring skyrmions as an interpolated map, which highlights the positions of the domain boundaries. The black rectangle serves as reference marker for the following analysis. **b** Considering every $\Delta\alpha > 10°$ as a DB, the grayscale map yields a probability map of hosting a DB during the 1000 frames of the 62.5 ms video of lattice formation.

order – either separately or even in combined approaches. Here, we use magnetic field oscillations as they act isotropic and since we can apply the excitation on the fly within the relevant field range and time scale. In contrast, additional patterning or devices are necessary for the other techniques.

During the driving by the oscillating field, the system is expected to exhibit non-equilibrium properties. Unlike for superconducting vortices however, the skyrmion order persists even when the driving mechanism (i.e., the field oscillations causing the depinning) is switched off. That is, after switching off the driving by the oscillating field, an equilibrium lattice with significantly enhanced order remains. However, there is a threshold for this mechanism: at large amplitudes (here, $A > 180$ μT), the increased diffusivity—similar to an elevated effective temperature—appears sufficient to counteract lattice stability, leading to a decreased order again. We further show that the diffusion coefficient $D$ of the skyrmions increases monotonously with $A$. At the same time, larger amplitudes cause increasing annihilation events and decrease the number of present skyrmions $N$ (Fig. 1c). The time evolution of all shown parameters as well as the correlation function $G_6$ are shown in Supplementary Fig. 1. Additionally, we characterize the effect in a similar scan of varying amplitude at $f = 500$ Hz as well as in a frequency scan at fixed amplitude in Supplementary Fig. 2. Here, we use the frequency of 100 Hz as it leads to a maximum enhancement of the skyrmion diffusion in this system.

Annihilations are a result of skyrmions being destabilized in the field oscillation. This effect has previously been observed for isolated skyrmions[21]. In a dense lattice, it becomes pronounced already for smaller amplitudes $A$ due to the strong skyrmion-skyrmion interactions present[14]. Therefore, especially skyrmions in “overpopulated” regions tend to annihilate. Note that the packing fraction—i.e., the area occupied by skyrmions—is predominantly set by the magnetic hysteresis. Therefore, skyrmion annihilations do typically not lead to a reduction of the packing fraction but to an increased size of the remaining skyrmions. Consequently, annihilations may solve space conflicts of individual skyrmions in “overpopulated” regions, which can facilitate the ordering process—provided the overall skyrmion density remains sufficient to stabilize an ordered lattice, as seen here. If too many skyrmions annihilate, the remaining skyrmions expand into stripes.

We note that despite the effective reduction of pinning effects due to the oscillating field[21], remaining effects of the non-flat energy landscape are still present[20]. As pinning effects have been shown to play an important role for breaking QLRO[11,12,27,40], we now analyze the role of the non-flat energy landscape during the lattice formation. This effective energy landscape could previously be determined directly from the occurrence probability of the skyrmions as a result of the spatially inhomogeneous skyrmion-material interaction[20,26]. However, in dense systems such as in a skyrmion lattice, skyrmion-skyrmion interactions between the particles themselves play an important role[14,16]. Those interactions also affect the occurrence distribution of skyrmions as illustrated in Supplementary Fig. 3, for instance when a skyrmion is confined by its lattice neighbors. Hence, the effects of the skyrmion-skyrmion interaction and the pining potential cannot be decoupled, and a direct determination of the energy landscape remains inaccessible in this regime. Instead, we access the effect of pinning indirectly though the resulting lattice properties in the following.

### Pinning effects confine lattice domains geometrically

As the non-flat energy landscape hampers QLRO and instead favors a “polycrystalline” structure with multiple domains, we first analyze the occurrence of lattice domains and domain boundaries. The orientation $\alpha$ is uniform within a lattice domain but changes at a DB. Therefore, we determine the orientation change $\Delta\alpha$ between neighboring skyrmions. We assign to every connection between nearest neighbors in the skyrmion lattice the absolute value of the change in orientation and interpolate these values onto the pixel grid. We show the determined $\Delta\alpha$ as background in Fig. 2a. For visualization, we overlay the corresponding orientation data from Fig. 1b as scattered dots, so that the orientation change $\Delta\alpha$ between the lattice domains (i.e., clusters of similar color) becomes clearly visible. The existence of a DB, however, is independent of the exact value of $\Delta\alpha$ – as long as it exceeds the typical fluctuations of a few degrees within a domain. Thus, we define all orientation changes $\Delta\alpha$ above a threshold of 10° as DB.

In Fig. 2b, we present the probability of hosting a DB for each pixel within 62.5 s (1000 frames) after nucleation. The domains fluctuate, rearrange, and grow from $t_0$ to $t_1$ (snapshots in Fig. 1a, b, respectively). Despite the fluctuations due to the diffusive dynamics, the DB occurrence probability map in Fig. 2b reveals significant spatial variations: some regions consistently host DBs, while others remain firmly inside lattice domains—suggesting an effective pinning of DBs.

To examine the ordering details on a local scale, we now zoom into the area marked by a black rectangle in Fig. 2. For this region of interest, we recall in Fig. 3a–c both the skyrmion positions (from Fig. 2a) and the DB occurrence probability map (from Fig. 2b), for reference. We color the occurring skyrmions by their local lattice orientation $\alpha$ (Fig. 3a), local order parameter $|\psi_6|$ (Fig. 3b) and number of lattice neighbors (Fig. 3c), respectively. The small red dots mark where the absolute orientation change $\Delta\alpha$

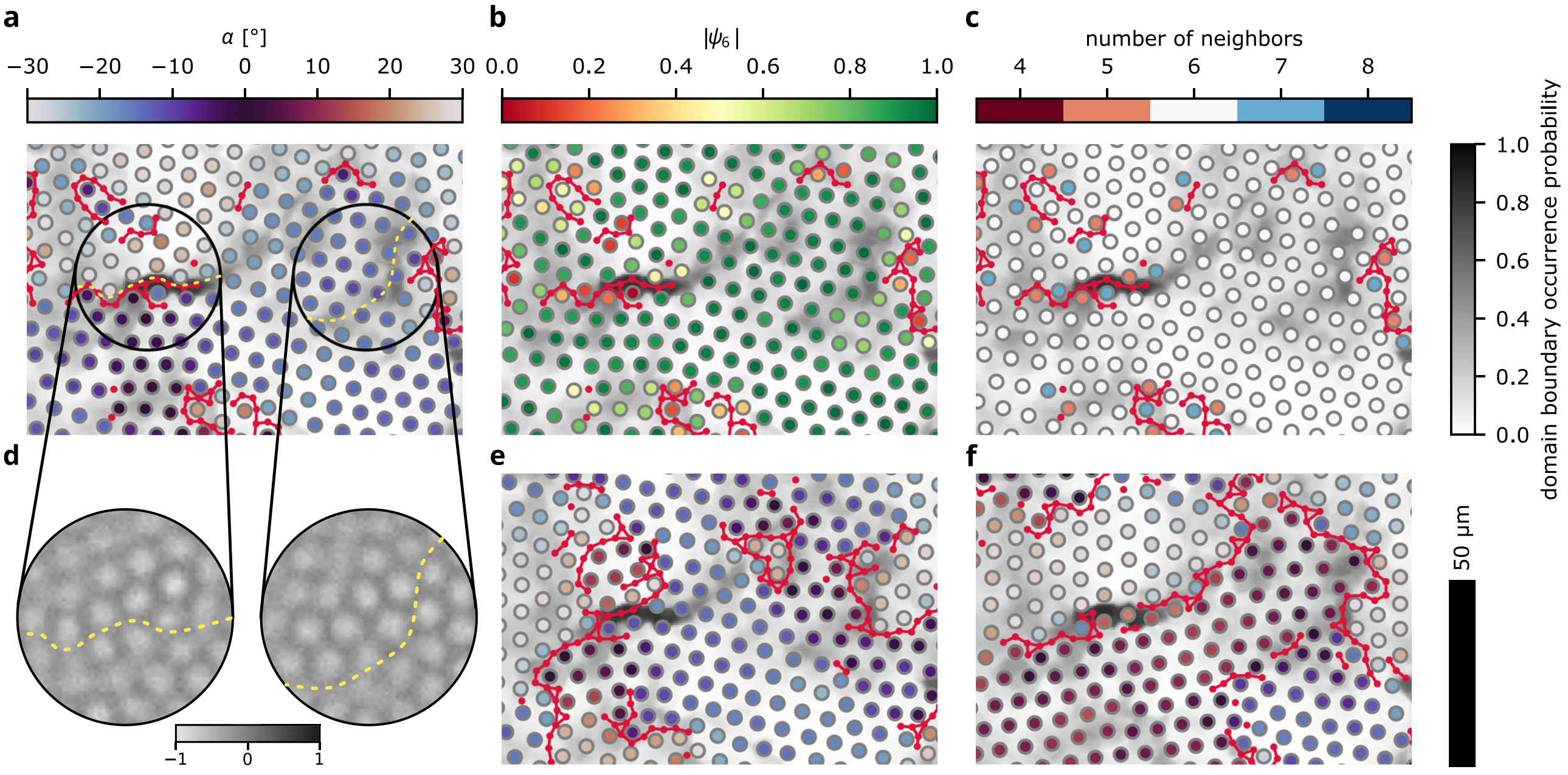

**Fig. 3 | Pinning effects induce lattice domain boundaries.** Onto the DB occurrence probability of the rectangle as greyscale background, we superimpose the **a** local orientation, **b** $|\psi_6|$ and **c** number of lattice neighbors for every skyrmion as dots using the data from Fig. 2a for a more detailed visualization. Red dots denote the occurrences of $\Delta\alpha > 10°$ between neighboring skyrmions and the red lines mark the corresponding domain boundaries. As the orientation changes along such boundaries, the local order $|\psi_6|$ is reduced and lines of topological defects occur. **d** Average Kerr contrast for three different Kerr videos of 1 min length in the marked circular areas. The yellow dashed line serves as marker to map positions of high domain wall occurrence probability from (**a**). Despite the three different nucleations, we find significant average contrast due to skyrmions at the positions where we have found high probability of a DB, thus indicating pinning effects. Contrarily, different configurations arise elsewhere as seen by the blurred contrast. **e**, **f** Plotting local orientation $\alpha$ of the two different nucleations (snapshots 60 s after nucleation) as colored dots onto the grayscale map of $\Delta\alpha$ from (**a**–**c**), we find again that domain boundaries of $\Delta\alpha > 10°$ (red lines) overlap with the DB occurrence probability from before, corroborating that pinning plays an important role for the occurrence of domain boundaries.

between two skyrmions exceeds 10°, contributing to a DB. Neighboring red dots are further connected by red lines to illustrate the DB contour. The marked red boundaries align well with positions where $\alpha$ changes in Fig. 3a, validating our DB identification. Since the orientational order is disrupted locally along a DB, we see in Fig. 3b that the corresponding local order parameter $|\psi_6|$ is reduced at these positions. This disruption also gives rise to topological defects, characterized by the number of lattice neighbors differing from the perfect order with six neighbors.

Next we probe the reproducibility of the skyrmion boundary formation. Since skyrmion pinning effects are a local material property, they are independent of the specific skyrmion configuration resulting from a nucleation event. Therefore, we compare the skyrmion occurrences in the same area but from different nucleations to analyze the DB pinning effect. Firstly, we investigate the circular areas marked in Fig. 3a, where the DB occurrence probability is particularly high. In Fig. 3d, we average the Kerr intensity for these circles across three videos, each capturing the first 62.5 s (1000 frames) of a newly nucleated skyrmion lattice. Despite averaging over three nucleations (with respective independent lattice domain formations) and long observation times relative to the diffusion timescale, we observe significant contrast. This contrast indicates that skyrmions recur at identical positions over time and across re-nucleations. For reference, we add (to Fig. 3d) a dashed yellow contour along which the DB occurrence probability in Fig. 3a is particularly high. These regions of high DB occurrence probability align well with the recurring skyrmions. In contrast, the intensity is more diffuse in areas further away from the primary DB positions, suggesting more fluctuations there. Therefore, we conclude that pinning effects play a crucial role in the formation of DBs within the skyrmion lattice, which inhibit QLRO and favor a polycrystalline multi-domain arrangement.

To further support this conclusion, we analyze snapshots showing the local orientation $\alpha$ (scattered dots) and DBs (red) from the second and third nucleation in Fig. 3e, f, respectively. In both cases, we plot the DB occurrence probability map from the first nucleation as the background (i.e., from Fig. 2b) for reference. The DBs identified in these additional snapshots frequently align with those from the first nucleation. As these DBs recur at identical positions across different nucleations, our observations confirm that the pinning of individual skyrmions effectively pins DBs within the lattice. As a result, pinned DBs suppress domain rearrangement towards QLRO, instead acting as geometric confinement for lattice domains.

To understand our experimental findings, we perform corresponding Brownian dynamics Thiele model simulations to elucidate the role of the non-flat energy landscape for the formation of lattice domains and DBs.

## Simulation results

We simulate skyrmion lattices within the Thiele model using a typical repulsive skyrmion interaction potential[10,14] (see Methods). The simulations are conducted at a fixed density of 1.25, which places the system deep within the solid phase in the absence of an external potential landscape[10,17].

First, we use an experimentally derived skyrmion energy landscape[20] as the potential energy input (Supplementary Fig. 4), varying only the depth of the potential landscape while keeping the skyrmion density constant. As shown in Fig. 4a, introducing a non-flat energy landscape results in a multi-domain skyrmion lattice. Pinning reduces the average local order $\psi_6$, the orientational correlation length $\xi_6$ and the diffusion coefficient $D$ (Fig. 4a). As the depth of the potential landscape decreases, all three quantities—$\psi_6$, $\xi_6$ and $D$—increase, consistent with the experimental observations. However, beyond a certain point, further increases in lattice order lead to a decrease in $D$, as the system favors preserving order over maximizing diffusivity. In the experiment, these opposing effects cannot be disentangled, as field oscillations primarily enhance diffusivity[21].

In Fig. 4b, we show that strong pinning results in small domains extending over only a few particles, while weaker pinning allows for increasingly uniform lattice orientation $\alpha$. For a selected region of interest (Fig. 4c), we present the experimental energy landscape used in the simulation and the associated DBs. The landscape consists of energy variations on the scale of the skyrmion diameter, forming randomly distributed pinning sites separated by energy barriers of varying height on the order of few $k_BT$. By analyzing 80 simulation time units (STU) of an equilibration, we demonstrate that this heterogeneous energy landscape leads to a highly inhomogeneous probability for domain boundary occurrence. Some DBs appear to remain persistently pinned by the landscape.

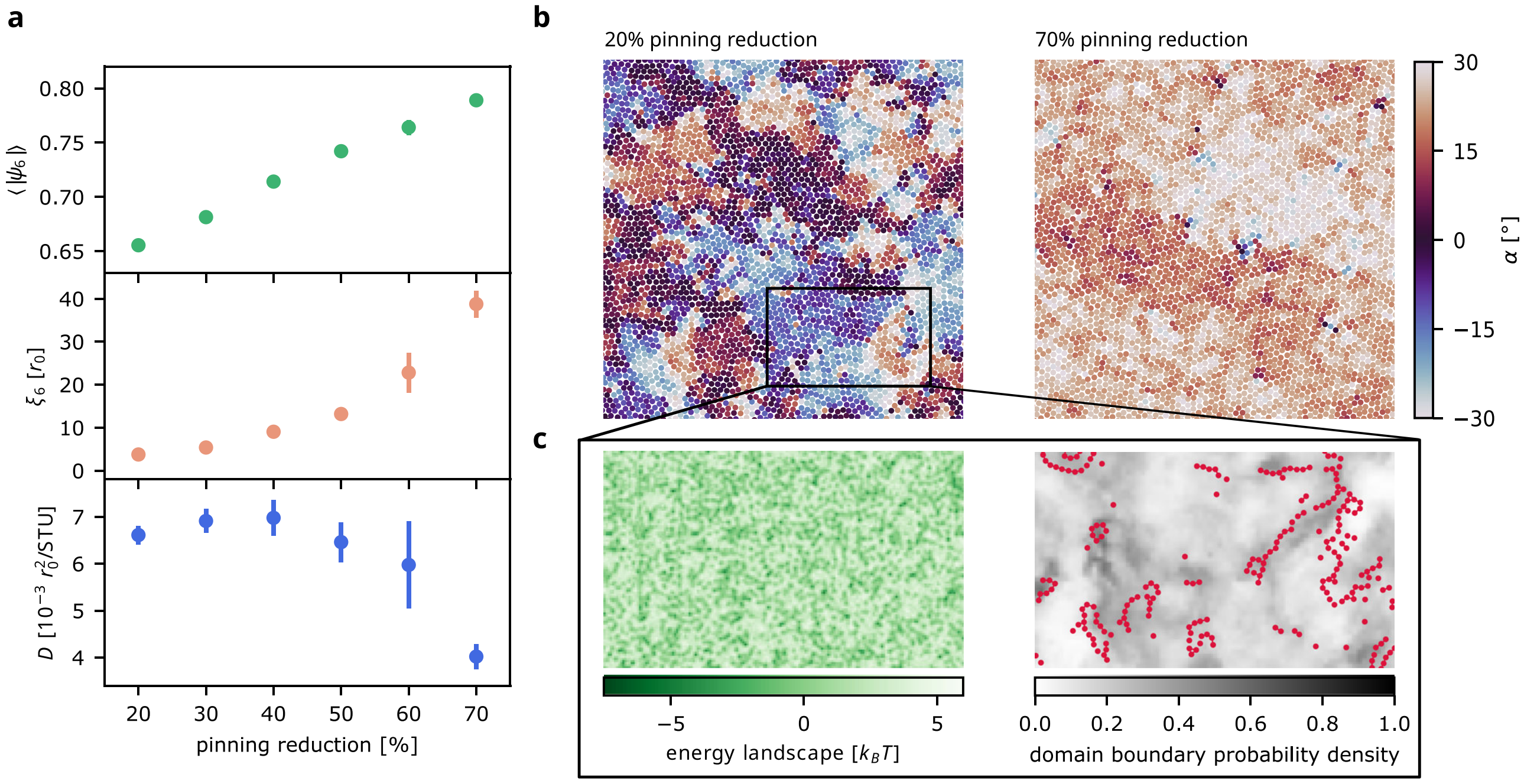

**Fig. 4 | Pinning leads to multi-domain states. a** We use an experimentally determined energy landscape in a Brownian dynamics simulation of 5300 particles at density 1.25, which is deep in the solid phase in the absence of a potential. The pinning leads to reduced local orientational order $\psi_6$, smaller correlation length $\xi_6$, and suppressed diffusivity. $D$ is given in terms of the nearest neighbor distance $r_0$ and the simulation time unit STU. When the pinning is reduced in the simulation, $\psi_6$, $\xi_6$, and $D$ increase. For increasing order, $D$ decreases again to preserve lattice order. Data points and error bars denote average and standard deviation, respectively, of the last 200 writeouts, averaged for four independent simulation runs each. **b** Local lattice orientation $\alpha$ for individual skyrmions (dots) at differently strong pinning. While stronger pinning with only 20% reduction of the energy landscape leads to polycrystalline lattice multi-domain state, reducing pinning by 70% results in an almost uniform $\alpha$. **c** For the rectangle highlighted in (**b**), we show the experimental energy landscape (left) and the occurring domain boundaries (right). The grayscale background is the domain boundary occurrence probability throughout the 800 writeouts in 80 STU. The red markers denote the domain boundaries identified in (**b**).

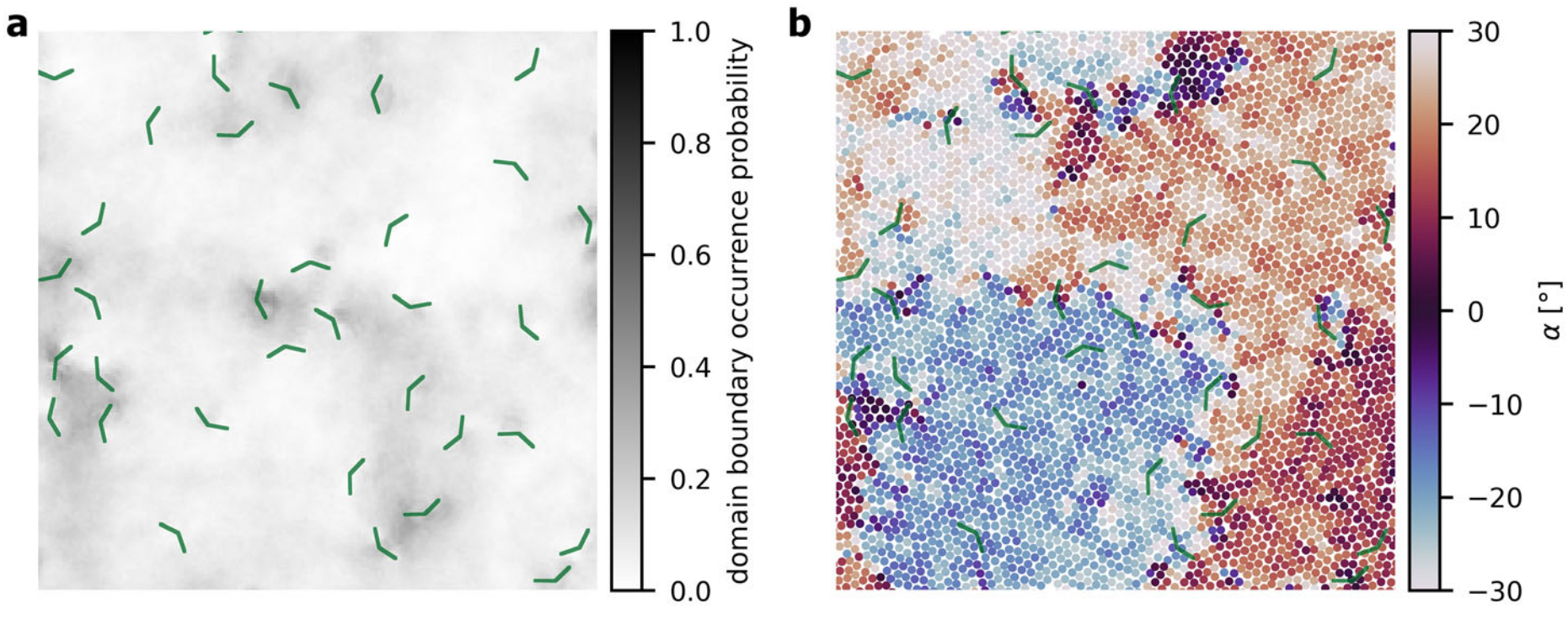

**Fig. 5 | Artificially designed pinning of lattice domain boundaries. a** In a simulation of 5000 particles at density 1.25, we introduce 40 randomly spread angle-shaped patterns (green) with an energy $-7.5\ k_BT$ with respect to the flat background potential. Throughout a simulation run, the domain boundary probability map remains inhomogeneous with regions of high occurrence probability around and between the angle patterns. **b** Snapshot of the local lattice orientation $\alpha$ per skyrmions after 100 STU. Distinct domains of different $\alpha$ form. The orientation changes in regions, where also the overall domain boundary occurrence probability is high. While some of the angle patterns (green) mark domain boundaries, others are located within a lattice domain.

To further understand the mechanism of DB pinning, we next introduce designed synthetic energy landscapes into the simulation. In Fig. 5, we employ a landscape composed of randomly distributed angle-shaped pinning features. Each feature spans three skyrmion spacings in both directions and has a depth of $-7.5\ k_BT$ relative to the flat background. The opening angle is set to 135°, which is deliberately chosen to be incommensurate with the hexagonal lattice. The background of Fig. 5a shows the DB occurrence probability over 80 STU, revealing a clearly inhomogeneous distribution. This minimal pattern suffices to create distinct lattice domains and to effectively pin DBs.

Figure 5b shows a snapshot of the local lattice orientation $\alpha$ per skyrmion after 80 STU. At several locations, $\alpha$ changes sharply at the angle-shaped pinning sites, which mark and anchor DBs. However, these angles can also be located within a single domain, merely inducing local distortions.

While domain configurations fluctuate, DBs tend to connect the angular pinning features, though the specific features involved vary over time.

To explore the influence of different pinning geometries, we compare several synthetic patterns. In the case of linear pinning features (Supplementary Fig. 5), domain fragmentation along the pinning lines is unfavorable, as the lines are commensurate with the lattice. Instead, the lines tend to stabilize the interior of a domain and confine DBs to the interstitial regions. Conversely, a landscape consisting of overlapping sinusoidal lines (Supplementary Fig. 6)—which are incommensurate with the hexagonal lattice—favors domain fragmentation. Since the sinusoidal lines are curved, a single domain covering the sine is unfavorable. Instead, breaking of the domains is induced. However, the continuous curvature does not anchor one fixed DB but instead causes small domains and DBs in its vicinity.

In the experimental system, the material's energy landscape is random in nature[21,26] and likely incommensurate with the hexagonal skyrmion lattice, as shown in Fig. 3d. Additionally, the observed pinning features predominantly exhibit a local character, resembled by the angle-shaped patterns used in the simulation (Fig. 5).

## Conclusions

The non-flat skyrmion energy landscape in magnetic thin films has been known to be the key inhibitor of quasi-long-range order (QLRO) of skyrmion lattices[11,12]. In this work, we demonstrate how skyrmion pinning effects in a non-flat energy landscape cause the local stabilization of lattice domain boundaries. The effectively pinned domain boundaries delineate and confine the polycrystalline lattice domains, breaking QRLO. As we can effectively tune the energy landscape and reduce pinning effects by magnetic field oscillations[21], we can significantly increase the size of the lattice domains, leading to enhanced average local order. We therefore provide a key enabler to access intrinsic properties and dynamics of skyrmion lattices[10] and their fundamental 2D phase behavior—within the limit of domains as well as towards single crystals of true QLRO in the future. The oscillating fields furthermore allow us to study both equilibrium and non-equilibrium phase behavior. Different mechanisms like current-induced motion[26,37], local heating[38] or surface acoustic waves[39] may also depin[36] and increase lattice domains in an alternative or additional approach. We reproduce our experimental results by quasi-particle simulations and we show that local modifications of the energy landscape[41,42] in order to deterministically pin single skyrmions at their ideal lattice sites pose an option to stabilize artificial lattices on larger scales—in a reverse engineering approach to access lattice properties beyond the current observations in systems with a natural energy landscape.

## Methods

### Magnetic multilayer material

The magnetic thin film stack is deposited using DC/RF magnetron sputtering in a *Singulus Rotaris* system under a base pressure of $3 \times 10^{-8}$ mbar. We use a stack composition of Ta(5 nm)/$Co_{20}Fe_{60}B_{20}$(0.9 nm)/Ta(0.08 nm)/MgO(2 nm)/$HfO_2$(3 nm) (with layer thicknesses in nanometers, accurate to within 0.01 nm). The perpendicular magnetic anisotropy (PMA, at the $Co_{20}Fe_{60}B_{20}$/MgO interface) and the interfacial Dzyaloshinskii-Moriya interaction (DMI, primarily at the Ta/$Co_{20}Fe_{60}B_{20}$ interface)[43,44] are key ingredients for the stabilization of magnetic skyrmions in our thin film. Thereby, the dusting layer of Ta(0.08 nm) optimizes the material for low pinning and skyrmion lattice formation by fine-tuning the balance between DMI and PMA[18,45]. The non-trivial topology of our magnetic skyrmion structures is confirmed by driving skyrmions using spin-orbit-torques as well as by micromagnetic simulations[18–20,46]. We show the magnetic hysteresis curve for an out-of-plane (OOP) field cycle loop in Supplementary Fig. 7.

### Experimental setup

To image magnetic skyrmions in the multilayer compound, we operate a commercially available Kerr microscope manufactured by *evico magnetics GmbH* in polar mode, resulting in grayscale contrast of the OOP magnetization. With a blue LED light source and a CCD camera, we reach a resolution of 200–400 nm and 62.5 ms (16 fps) in real-time and -space. We control and monitor the sample temperature with a Peltier element and a Pt100 with a precision of 0.1 K, respectively[18,20]. Enclosing the entire microscope in a thermally stabilized flow box further improves temperature stability. Electromagnetic coils provide independent magnetic fields in in-plane(IP) and OOP direction. The OOP field coil is custom-made to allow field control with sub-μT precision to ensure stable measurement conditions for the skyrmion system.

While a magnetic OOP offset field is required to stabilize skyrmions, a saturating IP field pulse is used to nucleate skyrmions while keeping the temperature constant. We control the density of the nucleated skyrmions by the OOP offset and IP pulsed field[20,47,48]. Under the varying oscillating fields described in the main text, we take videos of 62.5 s length (1000 frames) and detect skyrmions by the *trackpy* Python package[34], which is based on 2D Gaussian kernel fitting. Due to the skyrmions being a collective spin ensemble spanning over many pixels, the detection works with sub-pixel precision[20]. Except for the geometric patterned confinements, the magnetic film exceeds the field of view by more than one order of magnitude and is therefore considered as continuous, where we neglect boundary effects.

### Lattice order and diffusion analysis

From the detected skyrmion positions, we apply a Voronoi tessellation[35] to determine lattice neighbors. From the Voronoi grid, we determine the complex local order parameter $\psi_6$ for every skyrmion. The local lattice orientation directly follows as $\alpha = \arg(\psi_6)/6$[12]. We then calculate the orientational correlation function $G_6$ as a function of distance $r$ for every frame[10]. As no quasi-long-range order (QLRO) is present due to multi-domain structures[32], we determine the decay of the correlation function as an exponential $\propto \exp(-r/\xi_6)$ with a correlation length $\xi_6$[3,4], which we use as a measure for the lattice domain size. As an additional result of the Voronoi tessellation, we directly gain information about topological lattice defects[1,2]: every skyrmion with a number of lattice neighbors $N_{nn}$ different from 6 (which is the case in a purely hexagonal lattice structure) is a topological defect.

To evaluate the diffusivity of the skyrmions, we use *trackpy*[34] to link the skyrmion coordinates $\mathbf{r}$ from the single frames at time $t$ to trajectories. We then calculate the mean squared displacement

$$\mathrm{MSD}(t) = \left\langle \left[\mathbf{r}(t) - \mathbf{r}\left(t_0\right)\right]^2 \right\rangle = 2dDt$$

with respect to a start time $t_0$ by averaging over all skyrmion trajectories as indicated by the angled brackets[18,21]. Since the MSD is related to the diffusion coefficient $D$ via the time $t$ and the system dimensionality $d = 2$, we extract $D$ as a linear fit over a 1 s interval. The accessible fit region is limited in the experiment as it required reliable trajectory linking over the full fit window. We can use any time $t_0$ as starting time to evaluate the time evolution of the diffusion coefficient[10].

The occurrence probability $p$ of skyrmions at low skyrmion densities can furthermore be used to determine the energy landscape. Following a well-established procedure[20,26], we determine the energy landscape in units of $k_BT$ as $-\ln(p)$. The sample used for the lattice measurements features only high skyrmion densities at the measurement temperature. Reducing temperature allows also lower densities, but increases the skyrmion size (which is relevant for the energy landscape) and suppresses the relevant thermal skyrmion diffusion. Since sparse (and thermally diffusing) skyrmions are required in this procedure to minimize contributions from skyrmion-skyrmion interactions, we use a different, similar sample (Ta(5 nm)/$Co_{20}Fe_{60}B_{20}$(0.9 nm)/Ta(0.09 nm)/MgO(2 nm)/Ta(5 nm) measured at 316 K) to obtain a realistic energy landscape for the simulations.

### Brownian dynamics simulation

We use Brownian dynamics simulations in the Thiele model[49] to simulate thermally diffusing skyrmions. Sets $\{\mathbf{r}\}$ of skyrmions at positions $\mathbf{r}$ positions with velocity $\mathbf{v}$, the damping $\gamma$ (effective skyrmion damping for Brownian Dynamics simulation, not the Gilbert damping) and satisfy the equation of motion[14,26]

$$-\gamma\mathbf{v} - G_{\mathrm{rel}}\gamma\mathbf{e}_z \times \mathbf{v} + \mathbf{F}_{\mathrm{therm}} + \mathbf{F}_{\mathrm{SkSk}}(\{\mathbf{r}\}) + \mathbf{F}_{\mathrm{pin}}(\mathbf{r}) = 0$$

while using $\gamma = 1$ in simulation units. Since the relative Magnus force strength $G_{\mathrm{rel}}$ is negligible in our system, we do not consider the term in the simulations. The dynamics is influenced by several forces $\mathbf{F}$: the thermal white noise $\mathbf{F}_{\mathrm{therm}}$ (fulfilling the fluctuation-dissipation), skyrmion-skyrmion interactions $\mathbf{F}_{\mathrm{SkSk}}$, and pinning forces $\mathbf{F}_{\mathrm{pin}}$ originating from a non-flat energy landscape. Based on previous experiments[10,14], we use a skyrmion-skyrmion interaction potential $V(r) = r^{-8}$ with a cutoff distance of 3 simulation units. While theoretical studies[50–54] have shown the possibility to adjust the original Thiele equation by adding a mass term, we can confidently estimate the inertia of the skyrmion to be negligible in our system using previous determinations of mass[55] and damping[26].

The equation of motions are integrated using a Heun algorithm and periodic boundary conditions. Writeouts are performed every 10,000 simulation time steps, which corresponds to every 0.1 simulation time units (STU). The lattices are typically equilibrated after 20 STU (thus, 200 writeouts). The density of the skyrmions is determined as number of particles per squared simulation length unit. The system is initialized as a square lattice.

## Data availability

The data that support the findings of this study are available on Zenodo[56] under https://doi.org/10.5281/zenodo.17406183.

## Code availability

The computer code used for analyzing the experimental results as well as for generating and analyzing simulation data of this study is available from the corresponding author upon reasonable request.

## References

1. Kosterlitz, J. M. & Thouless, D. J. Long range order and metastability in two dimensional solids and superfluids. (Application of dislocation theory). *J. Phys. C* **5**, L124 (1972).
2. Kosterlitz, J. M. & Thouless, D. J. Ordering, metastability and phase transitions in two-dimensional systems. *J. Phys. C* **6**, 1181–1203 (1973).
3. Halperin, B. I. & Nelson, D. R. Theory of two-dimensional melting. *Phys. Rev. Lett.* **41**, 121–124 (1978).
4. Nelson, D. R. & Halperin, B. I. Dislocation-mediated melting in two dimensions. *Phys. Rev. B* **19**, 2457–2484 (1979).
5. Young, A. P. Melting and the vector Coulomb gas in two dimensions. *Phys. Rev. B* **19**, 1855–1866 (1979).
6. Bogdanov, A. & Hubert, A. Thermodynamically stable magnetic vortex states in magnetic crystals. *J. Magn. Magn. Mater.* **138**, 255–269 (1994).
7. Mühlbauer, S. et al. Skyrmion lattice in a chiral magnet. *Science* **323**, 915–919 (2009).
8. Jiang, W. et al. Skyrmions in magnetic multilayers. *Phys. Rep.* **704**, 1–49 (2017).
9. Huang, P. et al. Melting of a skyrmion lattice to a skyrmion liquid via a hexatic phase. *Nat. Nanotechnol.* **15**, 761–767 (2020).
10. Gruber, R. et al. Real-time observation of topological defect dynamics mediating two-dimensional skyrmion lattice melting. *Nat. Nanotechnol.* **20**, 1405–1411 (2025).
11. Meisenheimer, P. et al. Ordering of room-temperature magnetic skyrmions in a polar van der Waals magnet. *Nat. Commun.* **14**, 3744 (2023).
12. Zázvorka, J. et al. Skyrmion lattice phases in thin film multilayer. *Adv. Func. Mater.* **30**, 2004037 (2020).
13. Seshadri, R. & Westervelt, R. M. Statistical mechanics of magnetic bubble arrays. II. Observations of two-dimensional melting. *Phys. Rev. B* **46**, 5150–5161 (1992).
14. Ge, Y. et al. Constructing coarse-grained skyrmion potentials from experimental data with Iterative Boltzmann Inversion. *Commun. Phys.* **6**, 1–6 (2023).
15. Lin, S.-Z., Reichhardt, C., Batista, C. D. & Saxena, A. Particle model for skyrmions in metallic chiral magnets: dynamics, pinning, and creep. *Phys. Rev. B* **87**, 214419 (2013).
16. Jefremovas, E. M. et al. The role of magnetic dipolar interactions in skyrmion lattices. *Newton* **1**, 100036 (2025).
17. Kapfer, S. C. & Krauth, W. Two-dimensional melting: from liquid-hexatic coexistence to continuous transitions. *Phys. Rev. Lett.* **114**, 035702 (2015).
18. Zázvorka, J. et al. Thermal skyrmion diffusion used in a reshuffler device. *Nat. Nanotechnol.* **14**, 658–661 (2019).
19. Kerber, N. et al. Anisotropic skyrmion diffusion controlled by magnetic-field-induced symmetry breaking. *Phys. Rev. Appl.* **15**, 044029 (2021).
20. Gruber, R. et al. Skyrmion pinning energetics in thin film systems. *Nat. Commun.* **13**, 3144 (2022).
21. Gruber, R. et al. 300-times-increased diffusive skyrmion dynamics and effective pinning reduction by periodic field excitation. *Adv. Mater.* **35**, 2208922 (2023).
22. Zahn, K., Lenke, R. & Maret, G. Two-stage melting of paramagnetic colloidal crystals in two dimensions. *Phys. Rev. Lett.* **82**, 2721–2724 (1999).
23. Zahn, K. & Maret, G. Dynamic criteria for melting in two dimensions. *Phys. Rev. Lett.* **85**, 3656–3659 (2000).
24. Guillamón, I. et al. Direct observation of melting in a two-dimensional superconducting vortex lattice. *Nat. Phys.* **5**, 651–655 (2009).
25. Roy, I. et al. Melting of the vortex lattice through intermediate hexatic fluid in an α-MoGe thin film. *Phys. Rev. Lett.* **122**, 047001 (2019).
26. Brems, M. A. et al. Realizing quantitative quasiparticle modeling of skyrmion dynamics in arbitrary potentials. *Phys. Rev. Lett.* **134**, 046701 (2025).
27. Reichhardt, C., Reichhardt, C. J. O. & Milošević, M. V. Statics and dynamics of skyrmions interacting with disorder and nanostructures. *Rev. Mod. Phys.* **94**, 035005 (2022).
28. Iwasaki, J., Mochizuki, M. & Nagaosa, N. Universal current-velocity relation of skyrmion motion in chiral magnets. *Nat. Commun.* **4**, 1463 (2013).
29. Liu, Y.-H. & Li, Y.-Q. A mechanism to pin skyrmions in chiral magnets. *J. Phys.* **25**, 076005 (2013).
30. Navau, C., Del-Valle, N. & Sanchez, A. Interaction of isolated skyrmions with point and linear defects. *J. Magn. Magn. Mater.* **465**, 709–715 (2018).
31. Lima Fernandes, I., Bouaziz, J., Blügel, S. & Lounis, S. Universality of defect-skyrmion interaction profiles. *Nat. Commun.* **9**, 4395 (2018).
32. Deutschländer, S., Horn, T., Löwen, H., Maret, G. & Keim, P. Two-dimensional melting under quenched disorder. *Phys. Rev. Lett.* **111**, 098301 (2013).
33. Nelson, D. R. Reentrant melting in solid films with quenched random impurities. *Phys. Rev. B* **27**, 2902–2914 (1983).
34. Allan, D. B., Caswell, T., Keim, N. C., van der Wel, C. M. & Verweij, R. W. soft-matter/trackpy: v0.6.4. *Zenodo* https://doi.org/10.5281/zenodo.12708864 (2024).
35. Finney, J. L. & Bernal, J. D. Random packings and the structure of simple liquids. I. The geometry of random close packing. *Proc. R. Soc. A* **319**, 479–493 (1997).
36. Koshelev, A. E. & Vinokur, V. M. Dynamic melting of the vortex lattice. *Phys. Rev. Lett.* **73**, 3580–3583 (1994).

37. Woo, S. et al. Observation of room-temperature magnetic skyrmions and their current-driven dynamics in ultrathin metallic ferromagnets. *Nat. Mater.* **15**, 501–506 (2016).
38. Kern, L.-M. et al. Tailoring optical excitation to control magnetic skyrmion nucleation. *Phys. Rev. B* **106**, 054435 (2022).
39. Schwenke, P. et al. Ratchet motion of magnetic skyrmions driven by surface acoustic sawtooth waves. *Phys. Rev. B* **112**, 214409 (2025).
40. Reichhardt, C., Ray, D. & Reichhardt, C. J. O. Collective transport properties of driven skyrmions with random disorder. *Phys. Rev. Lett.* **114**, 217202 (2015).
41. Kern, L.-M. et al. Deterministic generation and guided motion of magnetic skyrmions by focused He+-ion irradiation. *Nano Lett.* **22**, 4028–4035 (2022).
42. Riddiford, L. J., Brock, J. A., Murawska, K., Hrabec, A. & Heyderman, L. J. Grayscale control of local magnetic properties with direct-write laser annealing. *Proc. SPIE PC13119, Spintronics XVII*, Vol. PC1311915 (SPIE, 2024).
43. Dzyaloshinsky, I. A thermodynamic theory of “weak” ferromagnetism of antiferromagnetics. *J. Phys. Chem. Solids* **4**, 241–255 (1958).
44. Moriya, T. Anisotropic superexchange interaction and weak ferromagnetism. *Phys. Rev.* **120**, 91–98 (1960).
45. Bhatnagar-Schöffmann, T. et al. Controlling interface anisotropy in CoFeB/MgO/HfO2 using dusting layers and magneto-ionic gating. *Appl. Phys. Lett.* **122**, 042402 (2023).
46. Rodrigues, D. R., Abanov, A., Sinova, J. & Everschor-Sitte, K. Effective description of domain wall strings. *Phys. Rev. B* **97**, 134414 (2018).
47. Zeissler, K. et al. Diameter-independent skyrmion Hall angle observed in chiral magnetic multilayers. *Nat. Commun.* **11**, 428 (2020).
48. Zivieri, R. et al. Configurational entropy of magnetic skyrmions as an ideal gas. *Phys. Rev. B* **99**, 174440 (2019).
49. Thiele, A. A. Steady-state motion of magnetic domains. *Phys. Rev. Lett.* **30**, 230–233 (1972).
50. Psaroudaki, C & Loss, D. Skyrmions driven by intrinsic magnons. *Phys. Rev. Lett*. **120**, 237203 (2018).
51. Reyes-Osorio, F. & Nikolić, B. K. Anisotropic mass of a magnetic skyrmion due to its interaction with conduction electrons: a Schwinger–Keldysh field theory approach. *J. Phys. Soc. Jpn.* **93**, 094707 (2024).
52. Psaroudaki, C., Hoffman, S., Klinovaja, J. & Loss, D. Quantum Dynamics of Skyrmions in Chiral Magnets. *Phys. Rev. X* **7**, 041045 (2017).
53. Pavlis, A. & Psaroudaki, C. Curvature-induced skyrmion mass. *Phys. Rev. Res*. **2**, 032058(R) (2020).
54. Wu, X. & Tchernyshyov, O. How a skyrmion can appear both massive and massless. *SciPost Phys.* **12**, 159 (2022).
55. Büttner, F. et al. Dynamics and inertia of skyrmionic spin structures. *Nat. Phys.* **11**, 225–228 (2015).
56. Gruber, R. et al. Source data - skyrmion lattice domain formation in a non-flat energy landscape. Data Set. *Zenodo* https://doi.org/10.5281/zenodo.17406183 (2025).

## Acknowledgements

This work was funded by the Deutsche Forschungsgemeinschaft (DFG, German Research Foundation) - SPP 2137 (project #403502522), TRR 173 Spin+X (projects A01, A12 and B02). The authors acknowledge funding from TopDyn. This project has received funding from the European Research Council (ERC) under the European Union’s Horizon 2020 research and innovation program (Grant No. 856538, project “3D MAGiC” and Grant No. 101070290, project “NIMFEIA”) and under the Marie Skłodowska-Curie grant agreements No. 860060 (“MagnEFi”) and No. 101119608 (“TOPO-COM”). The authors gratefully acknowledge the computing time granted on the supercomputer MOGON II and III at Johannes Gutenberg University Mainz as part of NHR South-West. M.A.B. was supported by a doctoral scholarship of the Studienstiftung des deutschen Volkes. E.M.J. acknowledges the Alexander von Humboldt Postdoctoral Fellowship. A.S. and M.K acknowledge support from the Norwegian Research Council through Grant No. 262633, Center of Excellence on Quantum Spintronics (QuSpin).

## Author contributions

R.G. performed the Kerr microscopy measurements and experimental data analysis with the help of T.S. J.R. and S.M.F. conducted the MD simulations with the help of M.A.B.; R.G., S.M.F and M.A.B analyzed the simulation data. F.K. and M.A.S. optimized and fabricated the multilayer stack. R.G. prepared the manuscript with the help of J.R., M.A.B., E.M.J. and S.K.; A.S., P.V. and M.K. guided and supervised the work. All authors have commented on the manuscript.

## Funding

Open Access funding enabled and organized by Projekt DEAL.

## Competing interests

The authors declare no competing interests.

## Additional information

**Supplementary information** The online version contains supplementary material available at https://doi.org/10.1038/s42005-025-02462-x.

**Correspondence** and requests for materials should be addressed to Mathias Kläui.

**Peer review information** *Communications Physics* thanks the anonymous reviewers for their contribution to the peer review of this work. A peer review file is available.